\documentclass[%
 aip,
 jmp,%
 amsmath,amssymb,
 reprint,%
]{revtex4-1}

\usepackage{graphicx}
\usepackage{dcolumn}
\usepackage{bm}
\usepackage[mathlines]{lineno}
\usepackage[usenames, dvipsnames]{color}
\usepackage{hyperref}
\begin{document}
\title{Electronic structure and Magneto-transport in MoS$_2$/Phosphorene van der Waals heterostructure}

\author{Sushant Kumar Behera}
 \affiliation{Advanced Functional Material Laboratory (AFML), Department of Physics, Tezpur University (Central University), Tezpur-784028, India}
\author{Pritam Deb*}%
 \email{pdeb@tezu.ernet.in (Corresponding Author)}

\date{\today}

\begin{abstract}
The time-dependent spin current mediated spin transfer torque behaviour has been investigated via scattering formalism within density functional theory framework supported by Green's function. Quantum magnetotransport characteristics 
have been revealed in a model semiconducting MoS$_2$/phosphorene van der Waals heterostructure. The dynamics of spin current channelized heterolayer transport has been studied with rotational variation in magnetization angle. It is 
observed that the time-dependent spin transport torque remains invariant irrespective of magnetization angle direction. The polarized spin-current is persistent with the external magnetic field for potential applicability towards spintronics.   
\end{abstract}

\maketitle

\section{Introduction}
Many interesting two dimensional (2D) materials, (like, h-BN, MoS$_2$, monolayer black phosphorous, etc.), are studied \cite{1, 2} following the research 
expansion on 2D materials. Phosphorene (P) was identified in this class as a relatively new material and grown experimentally under high pressure and temperature \cite{3, 4}. The narrow band gap ($~\sim$1.6 eV at monolayer) of 
Phosphorene is the primary reason for not being suitable for application in electronic devices \cite{5, 6}. In this aspect, molybdenum disulfide (MoS$_2$), an indirect bandgap (\,$~\sim$1.23 eV at monolayer) van der Waals (vdW) 
semiconductor TMDC, is considered to design 2D vdW heterostructure of MoS$_2$-P with minimum lattice mismatch (\,$~\sim$1$\%$) \cite{7,8,9}.  In recent times, 2D vertical heterostructures, like graphene-MoS$_2$ \cite{10}, 
graphene-h-BN \cite{11}, graphene-phosphorene \cite{12} and lateral heterostructures, like TMDC-TMDC \cite{13} for the vdW heterojunction, are modelled to investigate many novel phenomena, such as Hofstadter’s butterfly 
spectrum \cite{14, 15}, 
strongly bound exciton \cite{16} and spin valley polarization \cite{17} and transport (electronic or magnetic) behaviour. \\

Magneto-transport behaviour in confined dimension (known as Quantum magneto-transport) provides the realization 
about the channel transport 
properties of vdW 2D heterostructure. Besides, electronic angular-momentum helps in converting the electronic spins by transfer mechanism of the host heterostructure system. This process couples with the induced 
magnetization to generate the time-dependent phenomena ({\it i.e.} spin-transfer torque, STT) guided by spin-current \cite{18}. Moreover, conversion between spin to charge  
and {\it vice versa} generally occurs at the quantum regime close to the interface formed between two different surfaces of various electronic proeprties. A new dimension is added by exploring the channel transport 
properties of time varying spin-current in presence of external magnetic field (i.e. magneto-transport property), when vdW layers are integrated to form magnetic tunnel junctions (MTJs) \cite{19,20,21,22}.   \\

This new discovery has expanded to control the magnetization of magnetic materials with device applications \cite{23,24,25}. Furthermore, 
tunnel magnetoresistance is another important parameter which occurs at ferromagnetic-channel-metallic heterostructure. Here, DFT simulations are used combining scattering matrix formalism \cite{26} for designing 
our vdW MoS$_2$-P heterostructure system to investigate the time varying charge and spin transport. Non-equilibrium Green’s function (NEGF) is implemented sideways to DFT calculation for obtaining the exact behaviour of spin-current with subsequent magneto-transport property in this system. \\

\section{Methodology}
Density functional theory calculations are carried out via QE codes \cite{27} within LDA approximation \cite{28}. The vdW-DF scheme is used to include the van der Waals interaction. A $9\times9\times1$ k-points Monkhorst-Pack scheme is taken in the calculation at cutoff energy of 540 eV. Structural stability has been achieved when Hellmann-Feynman force limits within 10$^-3$ eV/$\AA{}$ per atom of the supercell. 12 $\AA{}$ is kept blank along z-axis to nullify the interaction among periodic images. $27\times27\times1$ k-point mesh is taken for DOS calculation \cite{29}. \\


\section{Results and Discussion}
We have calculated the magnetotransport behaviour of vdW-HS system consisting of phosphorene and MoS$_2$ monolayers as spin-valve defined in the parallel conﬁguration. 
Optimized geometry of the model vdW-HS is shown in Figure 1. In this aspect, the electronic density of states has been calculated for the bilayer system with respect to vacuum layer (shown in Supplementary Figure 1). 
Specially, the surface 
monolayer sheets are taken to realize the effective distribution of electron clouds in the heterostructure system for efficient spin tunneling. It is observed that the vacuum level is directly proportional to spin tunneling with 
increasing the states in conduction band for second vacuum level. Enhancement of states in vacuum level supports free movement of electronic spins and redistribution of electron clouds which directly corroborates improved tunneling 
and conductivity. Nature of STT of ferromagnetic region is gained in the heterostructure system. The trilayer MTJ is modelled here with two metallic 
surfaces (Fe and Pt) distincted by a semi-conducting nonmetallic vdW heterostructure (i.e. MoS$_2$-P) as spacer layer (shown in Figure 1 (c)). The system can be a potential candidate for exploiting STT effects (shown in Figure 1 (d)) and 
spintronics device applications. It is clear from the DOS pattern that the spin-polarized electrons reach to second layer accumulated at the interface.    \\

\begin{figure}[th!]     
\centering           
\includegraphics[width=8cm,height=5.5cm]{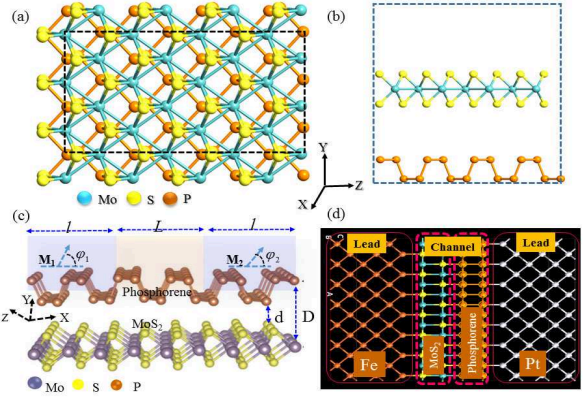}
\caption{\label{Fig:wide}Model heterostructure and high symmetry lines from (a) top view and (b) side view of the MoS$_2$-P. (c) Illustration of the model vdW heterojunction. Two metallic (Fe and Pt) regions having finite 
magnetization of different directional orientation have been sandwiched via tunnel region. (d) Scheme showing the magnetic tunnel junction (MTJ) formed by Fe/MoS$_2$P/Pt. Z axis is the path of electron flow.}
\end{figure} 

Here, total torque has been missed to get absorbed at the heterojunction interface (shown in Figure 2 (a) and (b)). Figure 2 (c) shows the STT effect of the MTJ junction. The torque components generate rapid oscillations. The penetration of the STT into the right ferromagnetic region in the vdW structure based heterojunction is similar to that in the ferromagnetic superconductor \cite{30} and  based 
spin valve of graphene \cite{31}.   \\

\begin{figure}[th!]     
\centering           
\includegraphics[width=8.5cm,height=3.5cm]{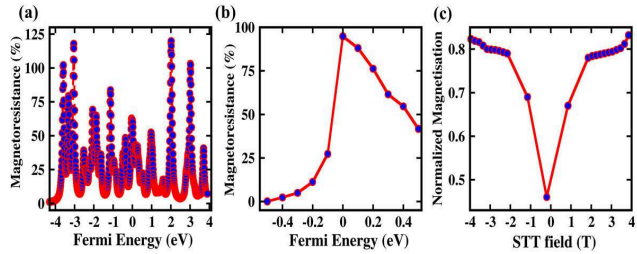}
\caption{\label{Fig:wide}Magnetoresistance profile of the heterolayer tunnel regions for MoS$_2$-P (a) at higher energy values and (b) at lower energy values. (c) Normalized magnetization strength has been shown as a 
function of STT field for the heterostructure system.}
\end{figure} 

Here, the theoretical model considers planar MTJ geometry of the vdW heterostructure system to interprete STT behaviour. Thus, it is needy to realize the STT change locally (time-dependent) for bilayer MTJs. 
To fulfill the criteria, we can impliment spin-polarized scanning tunneling microscopy (STM) to extract spin current behaviour which will help us to known transient transport trend locally \cite{32}. 
Current transport has been investigated from the model theory \cite{33,34,35} using the heterojunction. This correlation explains the complex wave function behaviour 
of the Bloch's equation for the system (shown in Figure 3 (a) and (b)).The average transmission coefﬁcients are gained in reciprocal space of the vdW heterostructure system 
(shown in Figure 3\,(c)). The coefficient of transmission for up and down electrons are about 0.054 and 0.0545 respectively close proximity to the Fermi level. The spin polarization is 
\,27 $\%$, presenting the majority electrons spin  dominating behaviour in such configuration. Figure 3 (d) presents the varying current of the heterostructure with parallel 
conﬁgurations under 10 mV.       \\

\begin{figure}[th!]     
\centering           
\includegraphics[width=8cm,height=5cm]{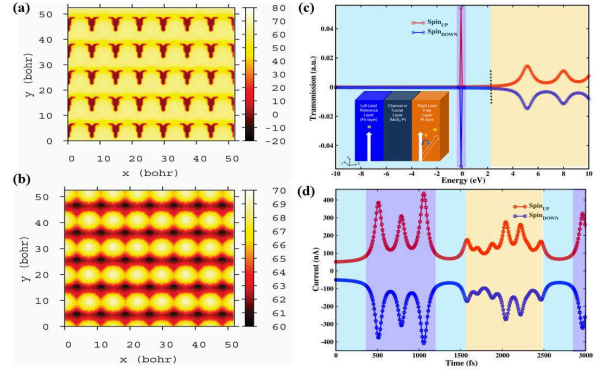}
\caption{\label{Fig:wide}Constant-current STM isosurface images of MoS$_2$-P heterolayer at (a) 10.0 mV and (b) -10.0 mV bias. (c) transmission-coefﬁcients for both up and down spins in the parallel 
conﬁgurations and (d) time-varying spin-polarized current of Fe/MoS$_2$-P/Pt layers with parallel conﬁguration under a finite pulse voltage of 10 mV. Scheme represents MTJ formed by Fe/MoS$_2$-P/Pt. Electrons flow through the z direction.  In (c) and (d), yellow, blue and violet color background present damping oscillation, consistent 
oscillation and steady state, respectively.}
\end{figure} 

The transient spin-current with STT have shown oscillatory behaviour supporting the current flow in the heterolayer channel. Figure 4 (a), 
(inset of Figure 4 (a)) shows STT of the channel heterostructure under direct current of 10 mV, 20 mV and 30 mV, respectively, for various magnetization orientation angles. 

\begin{figure}[th!]     
\centering           
\includegraphics[width=8.5cm,height=4cm]{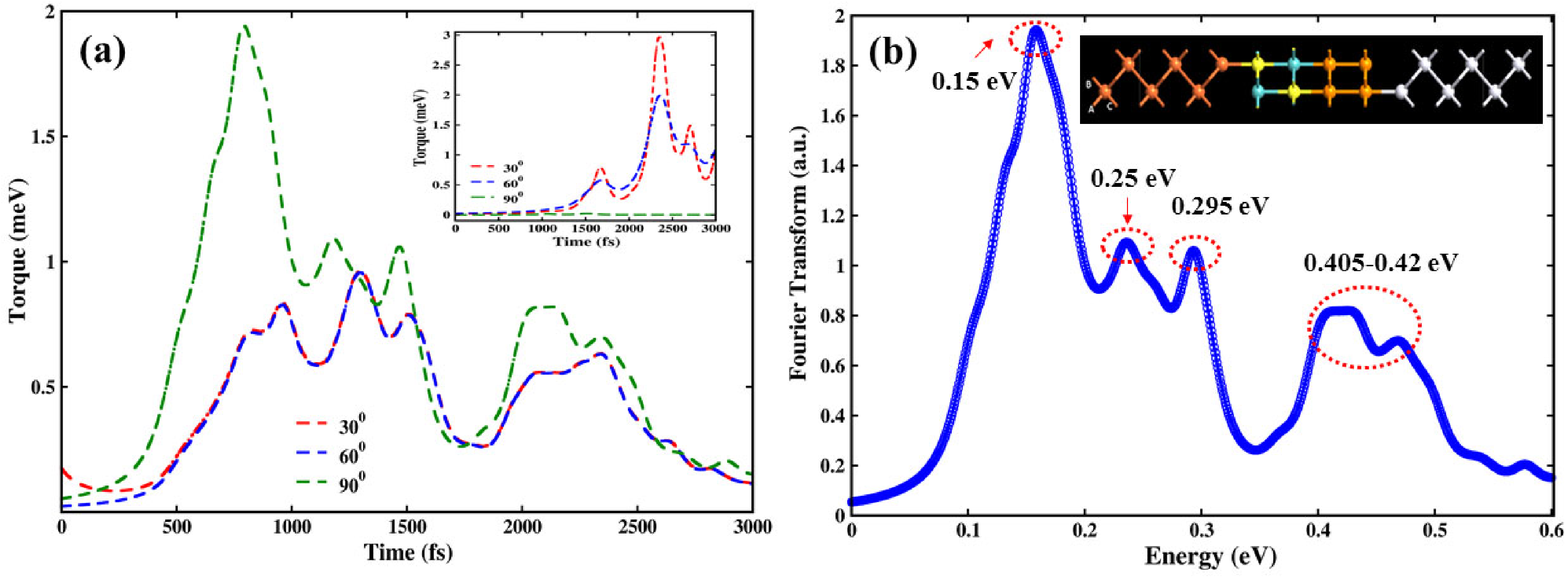}
\caption{\label{Fig:wide} Time-dependent spin-current induced STT of Fe/MoS$_2$/Pt layers for (a) 10 mV (inset for 20 mV). (b) time varying current in k-space for the 
Fe/MoS$_2$/Pt layers with $\theta$ = $\pi$/2 under a pulse voltage of 10 mV. Inset showing the MTJ formed using Fe, MoS$_2$ and Pt. }
\end{figure} 

We further demonstrate the STT infiltration into the right ferromagnetic region of the heterojunction. The STT is very sensitive to the chemical potential of the junction region as well as the exchange ﬁeld of the ferromagnetic 
region. The simulated STM images reveals detailed electronic structure of the heterostructure surface. The spin polarization of the conductance is about 27$\%$, dominating Fe/MoS$_2$-P/Pt layers in the parallel conﬁguration. 
Current-mediated magnetization switching in metallic spin valves and magnetic tunnel junctions (MTJs) is being important for possible applications in future spintronic/magnetic devices. Free-electron models have been employed earlier to study the electron transfer ({\it i.e.}\,STT) between two ferromagnets having a noncollinear magnetic alignment ignoring the local information on time-varying spin-current mediated channel transport which is mostly controlled by applied external magnetic field. In this regard, STM is an extremely useful method to provide local information on time varying spin-current behaviour in a wide variety of magnetic surfaces in heterostructure architecture. Thus, combining first principle calculations and spin-polarized STM simulation will be appropriate to investigate the detailed realization of the time-dependent spin-current and subsequent STT mediated magnetotransport behaviors in nanoscale semiconducting vdW heterostructure. This fact is addressed in this current chapter in MoS$_2$/phosphorene heterosystem.  
The time varying spin current and magneto-transport behavior are investigated in a model Fe/MoS$_2$-phosphorous/Pt vdW-HS system using first principle based simulations. It is clear STT modulated tunnel magneto-resistance behaviour plays key role to control quantum magnetotransport effects in 2D van der Waals heterojunctions. Electronic states show high value of charge accumulation in the heterojunction case supporting effective channel formation in the junction region. The spin current generated via tunnel magnetoresistance is directly proportional to the STT value indicating effective magnetization switching. Fast switching helps in channelizing high quantum of time-varying spin-current flow. Spin current shows damped oscillatory transport behavior with significant increment in the STT coefficient. These spin dynamics makes the heterostructure system a proximate platform in confined dimension towards spintronics application.

\section{Conclusion}
In conclusion, spin-transfer torque modulated tunnel magneto-resistance behaviour plays key role to control quantum magnetotransport effects in 2D van der Waals heterojunctions. Here, we investigate the time varying spin 
current and magnetotransport behavior in a model Fe/MoS$_2$-P/Pt vdW-HS system using first principle based simulations. Electronic states show high value of charge accumulation in the heterojunction case supporting effective 
channel formation in the junction region. The spin current generated via tunnel magnetoresistance is directly proportional to the STT value indicating effective magnetization switching. Fast switching helps in channelizing 
high quantum of time-varying spin-current flow. Spin current shows damped oscillatory transport behavior with an increment of 27$\%$ in the spin transfer torque coefficient. These spin dynamics makes the heterostructure system a 
proximate platform in confined dimension towards spintronics application. \\

\begin{acknowledgments}
S.K.B. acknowledges to Department of Science and Technology, DST, Government of India for INSPIRE Fellowship. The authors would like to thank Tezpur University for providing High Performing Cluster Computing (HPCC) facility.  \\
\end{acknowledgments}

\nocite{*}
\bibliography{manuscript}
\end{document}